\documentclass[prd,twocolumn,showpacs,floatfix,amsmath,amssymb,floatfix]{revtex4}
\usepackage{graphicx,color,dcolumn,booktabs,bm}
\usepackage{longtable,lscape}
\usepackage{txfonts}
\usepackage{overpic}
\usepackage{amssymb}
\usepackage{indentfirst}
\usepackage{feynmf}
\usepackage{slashed}
\usepackage{cases}
\usepackage{color}
\usepackage{multirow}
\usepackage{epstopdf}
\usepackage{graphicx,color,dcolumn,booktabs,bm}

\graphicspath{{Figures/}} %
\usepackage[colorlinks,
                      citecolor=blue,
                      anchorcolor=red,
                      menucolor=red,
                      linkcolor=red,
                      filecolor=red,
                      runcolor=red,
                      urlcolor=blue,
                      frenchlinks=red]{hyperref}
\begin{document}

\title{Production of $P_{cs}(4459)$ from $\Xi_b$ decay}
\author{Qi Wu$^{1}$}
\author{Dian-Yong Chen$^{1}$\footnote{Corresponding author}} \email{chendy@seu.edu.cn}
\author{Ran Ji$^{2}$}
\affiliation{
 $^{1}$ School of Physics, Southeast University,  Nanjing 210094, China\\
 $^{2}$ Research School of Physics, Australian National University, Canberra 2601, Australia}

\begin{abstract}
Inspired by the $P_{cs}(4459)$ reported by the LHCb collaboration recently, we investigate the $P_{cs}(4459)$ production from $\Xi_b$ decay in a molecular scenario using an effective Lagrangian approach. With different $J^P$ assignments to $P_{cs}(4459)$, the magnitude of branching fractions of $\Xi_b \to P_{cs}(4459) K$ is estimated, which is of the order of $10^{-4}$. Together with the decay properties of $P_{cs}(4459)$, the present estimations could be further testified by precise measurements and contribute to a better understanding of the molecular interpretations and the exploration of $J^P$ quantum numbers of $P_{cs}(4459)$.

\end{abstract}


\maketitle

\section{Introduction}
\label{sec:introduction}

The notion of pentaquark can date back to the birth of the quark model. The searches of pentaquark states become one of the most crucial inspections on the quark model. A lot of efforts have been made from both experimental and theoretical aspects and progresses have been achieved in recent decades (see recent reviews~\cite{Liu:2019zoy, Chen:2016qju, Guo:2017jvc} for more details). Among these achievements, the observations of $P_c$ states are undoubtedly one of the most important experimental breakthroughs, which was initially reported by the LHCb collaboration in 2015. The two pentaquark candidates, $P_c(4380)$ and $P_c(4450)$, were  observed in the $J/\psi p$ invariant mass distribution of $\Lambda_b \to K J/\psi p$ process~\cite{Aaij:2015tga}. And later in 2019, with more data samples, the LHCb collaboration updated their analysis of the $J/\psi p$ invariant mass distribution of $\Lambda_b \to K J/\psi p$ and discovered the split of structures $P_c(4380)$ and $P_c(4450)$ into three structures, namely $P_c(4312)$, $P_c(4440)$ and $P_c(4457)$ \cite{Aaij:2019vzc}.

Since these $P_c$ states were observed in the $J/\psi p$ invariant mass distribution, the most possible quark components are $c \bar{c} uud$, which indicates that these $P_c$ states can be good candidates of pentaquark states. In the pentaquark scenario, the spectroscopy and decay properties  have been investigated using the QCD sum rule \cite{Chen:2015moa,Wang:2015epa,Wang:2019got} and the constituent quark model \cite{Ortega:2016syt,Park:2017jbn,Weng:2019ynv,Zhu:2019iwm}. In the vicinity of $P_c$ states, there are abundant thresholds of a baryon and a meson, such as $\Sigma_c^{(\ast)} \bar{D}^{(\ast)}$, $\Lambda_c \bar{D}^\ast$, $\chi_{c1} p$, $\psi(2S) p$. In particular, the mass of $P_c(4312)$ is very close to the threshold of $\Sigma_c \bar{D}$, while $P_c(4440)$ and $P_c(4457)$ are close to the threshold of $\Sigma_c \bar{D}^{\ast}$. Considering the $S$ wave interaction, the possible $J^P$ quantum numbers of $\Sigma_c \bar{D}$ are $1/2^{-}$, while for $\Sigma_c \bar{D}^{\ast}$ system, the possible $J^P$ quantum numbers are $1/2^-$ and $3/2^-$, which illustrates that $P_c(4440)$ and $P_c(4457)$ could be assigned as a $\Sigma_c \bar{D}^\ast$ molecular state with different $J^P$ quantum numbers. Along this way, the mass spectrum \cite{Chen:2015loa,He:2015cea,Chen:2019bip,Liu:2019tjn,Azizi:2016dhy,Huang:2015uda,Chen:2019asm,He:2019ify,He:2019rva,Zhang:2019xtu}, decay properties \cite{Wang:2015qlf,Lu:2016nnt,Shen:2016tzq,Lin:2017mtz,Guo:2019fdo,Xiao:2019mvs,Wang:2019hyc,Xu:2019zme,Lin:2019qiv,Dong:2020nwk} and production behaviors \cite{Wang:2019krd,Wu:2019rog,Wang:2019dsi} of these $P_c$ states have been investigated in various methods, such as the QCD sum rule \cite{Azizi:2016dhy,Chen:2019bip,Wang:2019hyc,Xu:2019zme,Zhang:2019xtu}, the potential model \cite{Chen:2015loa,He:2015cea,Huang:2015uda,Chen:2019asm,Liu:2019tjn,He:2019ify,He:2019rva} and the effective Lagrangian approach \cite{Lu:2016nnt,Xiao:2019mvs,Wang:2019krd,Wu:2019rog,Wang:2019dsi,Lin:2019qiv}.

Inspired by the $P_c$ states, the authors in Refs. \cite{Hofmann:2005sw,Wu:2010vk,Cheng:2015cca,Anisovich:2015zqa,Wang:2015wsa,Feijoo:2015kts,Chen:2015sxa,Chen:2016ryt,Lu:2016roh,Xiao:2019gjd,Wang:2019nvm,
Park:2018oib,Shen:2020gpw,Zhang:2020cdi,Dong:2021juy} investigated the existence of the hidden-charm pentaquark states with strangeness, named as $P_{cs}$ states. Based on the SU(3) flavor symmetry, the authors in Ref. \cite{Cheng:2015cca} advised to search the $P_{cs}$ states in $\Xi_b\rightarrow P_\Sigma K$ and $\Omega_b\rightarrow P_\Xi K$ which may consist rates comparable to $\Lambda_b\rightarrow P_p K$. In Refs. \cite{Anisovich:2015zqa,Wang:2015wsa}, hidden-charm pentaquark states with strangeness were studied in the diquark-diquark-antiquark picture. The $J/\psi\Lambda$ invariant mass distribution was studied in reactions of $\Lambda_b\rightarrow J/\psi\eta\Lambda$ \cite{Feijoo:2015kts}, $\Lambda_b\rightarrow J/\psi K^0\Lambda$ \cite{Lu:2016roh} and $\Xi^-_b\rightarrow K^- J/\psi\Lambda$ \cite{Chen:2015sxa,Shen:2020gpw} to examine the existence of the hidden-charm pentaquark states with strangeness. In the molecular scenario, the prediction of hidden-charm pentaquark states with strangeness was provided in Refs. \cite{Chen:2016ryt,Xiao:2019gjd,Zhang:2020cdi}.

Experimentally, one predicted $P_{cs}$ state was reported very recently by the LHCb collaboration in the $J/\psi\Lambda$ invariant mass spectrum of the $\Xi^-_b\rightarrow J/\psi \Lambda K^-$ decays for the first time~\cite{Aaij:2020gdg}. The measured mass and width are, respectively,
\begin{eqnarray}
m_{P_{cs}} &=&	4458.8\pm2.9^{+4.7}_{-1.1} \ \mathrm{MeV}, \\
\Gamma_{P_{cs}} &=&	17.3\pm6.5^{+8.0}_{-.57} \ \mathrm{MeV}.
\end{eqnarray}
However, the $J^P$ quantum numbers of $P_{cs}(4459)^0$ were not determined.

It should be noticed that the mass of $P_{cs}(4459)^0$ is about 19 MeV below the threshold of $\Xi_c \bar{D}^\ast$, which naturally assigns $P_{cs}(4459)^0$ to a hadronic molecule state composed of $\Xi_c \bar{D}^\ast$ with possible $J^P$ quantum numbers $1/2^-$ or $3/2^-$. As a strange partner of $P_c$, the properties of $P_{cs}(4459)$ was investigated in molecular~\cite{Chen:2020uif,Peng:2020hql,Chen:2020opr,Chen:2020kco,Liu:2020hcv} and pentaquark \cite{Wang:2020eep} scenarios after the observation.  Along the way of $\Xi_c \bar{D}^\ast$ molecular scenario, we investigate the production mechanism of $\Xi^-_b\rightarrow P_{cs}(4459)^0 K^-$ using an effective Lagrangian approach. In such an estimation, we regard $P_{cs}(4459)^0$ as a $\Xi_c \bar{D}^\ast$ molecule state, while the probable conditions $J^P=1/2^-$ and $3/2^-$ are both considered. Similar to the case of $P_{c}$ state~\cite{Wu:2019rog, Xiao:2019mvs}, the present  estimation helps to test the molecular interpretations and verify the quantum number of $P_{cs}$ state in the molecule ansatz.

The present work is organized as the following. After the introduction, the formulae of the productions of  $\Xi_b \to P_{cs}(4459) K$  are shown, including the relevant effective Lagrangians and production amplitudes. Also, we exhibit our numerical estimation and some discussions on our results. Final, we give a short summary.

\section{The productions of $\Xi_b \to P_{cs} K$}
\label{sec:Sec2}
\begin{figure}[htb]
\begin{tabular}{ccc}
  \centering
  \includegraphics[width=2.8cm]{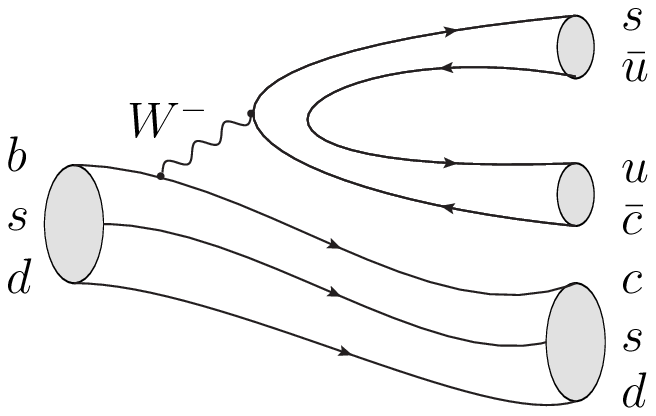}&
 \includegraphics[width=2.8cm]{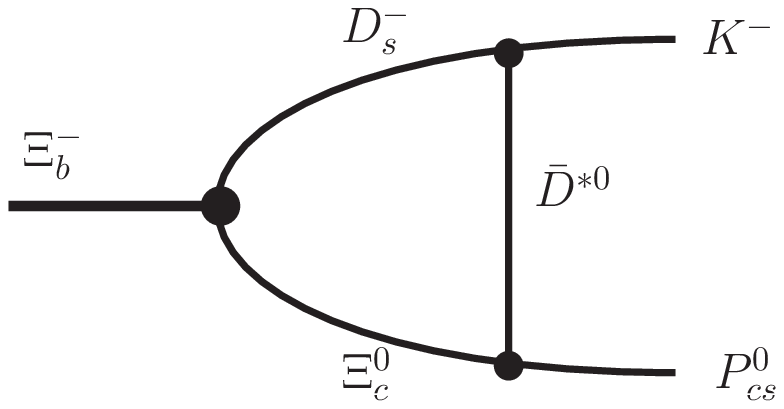}&
 \includegraphics[width=2.8cm]{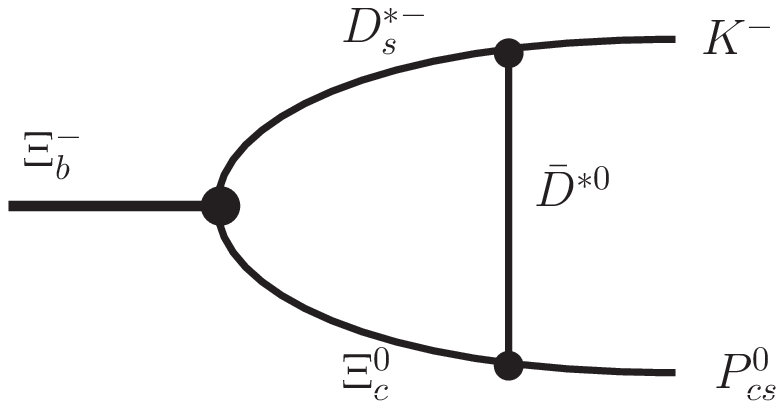}\\
 \\
 $(a)$ & $(b)$ & $(c)$ \\
 \end{tabular}
  \caption{Diagrams contributing to $\Xi_b\rightarrow P_{cs} K$ at the quark level (diagram(a)) and hadron level (diagrams(b)-(c)).}\label{Fig:Tri}
\end{figure}

The quark components of $\Xi_b$ are $bsd$, while the quarks in the final states are $ s\bar{u}(K)  csd (\Xi_c) \bar{c} u (\bar{D}^\ast)$. At the quark level, the process $\Xi_b \to P_{cs} K$ (Here and after, $P_{cs}$ refers to $P_{cs}(4459)$.)  occurs via the following subprocess from the phenomenological point of view: (1) The bottom quark transits to charm quark by emitting a $W^-$ boson which couples to $\bar{c} s$; (2) the $\bar{c} s$ and the $u\bar{u}$ pair created from vacuum transits to $K^-$ and $\bar{D}^\ast$, then $\bar{D}^\ast$ and $\Xi_c$ form a bound state, i.e., $P_{cs}$. For simplicity, $H_W$ and $H_T$ are used to denote the first and the second subprocess respectively. The production of $P_{cs}$ can be written as	$\langle K P_{cs} |H| \Xi_b \rangle =\langle K P_{cs} |H_W H_T| \Xi_b \rangle$. The direct estimations of this production process at quark level are much more difficult. In the present work, we qualitatively estimate this process at hadron level by inserting a complete basis formed by a meson and a baryon between $H_W$ and $H_T$, then the production process of $P_{cs}$ becomes,
\begin{eqnarray}
\langle K P_{cs} |H| \Xi_b \rangle= \sum_{B,M} \langle KP_{cs} |H_T| BM\rangle \langle BM|H_W|\Xi_b\rangle.
\end{eqnarray}
In principle, all the possible basis that can connect the initial $\Xi_b$ and $K P_{cs}$ should be considered. In the molecular scenario, the molecule should couple to its constituent hadrons more strongly than to other hadrons. Moreover, the loops composed by the ground states of the baryons and mesons are regarded as the major contributions as indicated in Refs.~\cite{Colangelo:2002mj,Colangelo:2003sa,Xiao:2016hoa,Chen:2016byt,Xiao:2018kfx, Chen:2015igx}. With such an assumption, the dominant diagrams contributing to $\Xi_b \to P_{cs} K$ at hadron level are presented in diagrams (b) and (c) in Fig.~\ref{Fig:Tri}.

\subsection{Effective Lagrangian}

In the present work, the diagrams at the hadron level are estimated by an effective Lagrangian approach.  In the molecular scenario, the $P_{cs}$ is considered to be molecular state composed of $\bar{D}^\ast \Xi_c$ with $J^P =1/2^-$ or $3/2^-$ and the simplest coupling of $P_{cs}$ with its components are~\cite{Zou:2002yy}
\begin{eqnarray}
\mathcal{L}^{1/2^-}_{p_{cs}\Xi_c \bar{D}^\ast} &=& g_{p_{cs}}^{1/2^-}\  \bar{\Xi}_c \gamma_5 \left(g_{\mu\nu}-\frac{p_{\mu}p_{ \nu}}{m^2}\right)\gamma^\nu P_{cs} \bar{D}^{\ast\mu},\nonumber\\
\mathcal{L}^{3/2^-}_{p_{cs}\Xi_c \bar{D}^\ast} &=& g_{p_{cs}}^{3/2^-}\ \bar{\Xi}_c P_{cs\mu} \bar{D}^{\ast\mu},\nonumber\\ \label{eq:3}
\end{eqnarray}
where $\mathcal{L}^{1/2^-}_{p_{cs}\Xi_c \bar{D}^\ast}$ and $\mathcal{L}^{3/2^-}_{p_{cs}\Xi_c \bar{D}^\ast}$ denote the Lagrangians for $J^P=1/2^-$ and $3/2^-$ respectively; $p$ and $m$ are the momentum and mass of $P_{cs}$, respectively.

The effective Lagrangians relevant to $\Xi_b \rightarrow \Xi_c D^{(\ast)}_s$ are~\cite{Cheng:1996cs}
\begin{eqnarray}
\mathcal{L}_{\Xi_b \Xi_c D_s} &=& i\bar{\Xi}_c \left(A+B\gamma_5\right)\Xi_b D_s,\nonumber\\
\mathcal{L}_{\Xi_b \Xi_c D^\ast_s} &=& D^{\ast\mu}_s \bar{\Xi}_c \left(A_1 \gamma_\mu \gamma_5+A_2 p_{2\mu}\gamma_5+B_1 \gamma_\mu+B_2 p_{2\mu} \right)\Xi_b,\nonumber\\
 \label{eq:1}
\end{eqnarray}
where $p_2$ is the momentum of $\Xi_c$. $A, \ B$, $A_1,\ A_2,\ B_1$ and $B_2$ are the combinations of the form factors $f_1$, $f_2$, $g_1$ and $g_2$, which are~\cite{Cheng:1996cs}
\begin{eqnarray}
  A &=& \lambda a_1 f_{D_s}\left(m-m_2\right) f_1,\nonumber\\
  B &=&\lambda a_1 f_{D_s} \left(m+m_2 \right) g_1,\nonumber\\
  A_1 &=& -\lambda a_1 f_{D^\ast_s}m_1\left[ g_1+g_2\left(m-m_2\right) \right],\nonumber\\
  A_2&=& -2\lambda a_1 f_{D^\ast_s}m_1 g_2,\nonumber\\
  B_1 &=& \lambda a_1 f_{D^\ast_s}m_1 \left[f_1-f_2\left(m+m_2\right)\right],\nonumber\\
   B_2&=& 2\lambda a_1 f_{D^\ast_s}m_1 f_2,
\end{eqnarray}
with $\lambda=\frac{G_F}{\sqrt{2}}V_{cb}V^\ast_{cs}$, and $a_1=1.07$~\cite{Li:2012cfa}. Here $m$, $m_1$ and $m_2$ are the masses of $\Xi_b$, $D^{(\ast)}_s$ and $\Xi_c$ respectively; $f^{(\ast)}_{D_s}$ is the decay constant of a charmed strange meson, which is estimated by a twisted-mass Lattice QCD~\cite{Carrasco:2014poa}. The form factors $f_i$ and $g_i$ (i=1,2) will be discussed in the following.

Following Refs.~\cite{Lin:1999ad, Oh:2000qr} and starting with SU(4) Lagrangian for pseudoscalar and vector meson, we can construct the effective Lagrangian relevant for $D_s^{(\ast)} D^\ast K$,
\begin{eqnarray}
\mathcal{L}_{KD_sD^{\ast}}&=&i g_{KD_sD^\ast}D^{\ast\mu}\left[\bar{D}_s\partial_\mu K -\left(\partial_\mu\bar{D}_s \right)K \right]+H.c.,\nonumber\\
\mathcal{L}_{KD^\ast_s D^{\ast}}&=&-g_{KD^\ast_s D^\ast}\varepsilon^{\mu\nu\alpha\beta} \left(\partial_\mu\bar{D}^\ast_\nu \partial_\alpha D^\ast_{s\beta}\bar{K}\nonumber+\partial_\mu D^\ast_{\nu}\partial_\alpha\bar{D}^\ast_{s\beta}K\right), \nonumber\\\label{eq:2}
\end{eqnarray}
where the coupling constants are $g_{KD_sD^\ast}=5.0$ and $g_{KD^\ast_s D^\ast}=7.0 \,\mathrm{GeV}^{-1}$ ~\cite{Lin:1999ad, Oh:2000qr}.

\subsection{Decay Amplitude}

With the above effective Lagrangians, we can obtain the amplitudes corresponding to the diagrams (b) and (c) in Fig.~\ref{Fig:Tri}. With the assumption that the $J^P$ quantum numbers of $P_{cs}(4459)$ are $\frac{1}{2}^-$, the decay amplitudes could be derived as follows:
\begin{eqnarray}
\mathcal{M}_b^{1/2}&=&i^3 \int\frac{d^4 q}{(2\pi)^4}\Big[g_{p_{cs}}^{1/2^-}\bar{u}(p_4)\gamma^\nu \gamma_5 \left(g_{\mu\nu}-\frac{p_{4\mu}p_{4\nu}}{m^2_4}\right)\Big]\nonumber\\
&&\times\Big(p_2\!\!\!\!\!\slash+m_2\Big)\Big[i(A+B\gamma_5)u(p)\Big] \Big[ig_{KD^\ast D_s}(ip_{3\alpha}+ip_{1\alpha}) \Big]  \nonumber\\
&&\times \frac{1}{p^2_1-m^2_1}\frac{1}{p^2_2-m^2_2}\frac{-g^{\mu\alpha}+q^\mu q^\alpha/m^2_E}{q^2-m^2_E}\mathcal{F}(q^2,m_{E}^2).\nonumber\\
\mathcal{M}_c^{1/2}&=&i^3 \int\frac{d^4 q}{(2\pi)^4}\Big [g_{p_{cs}}^{1/2^-}\bar{u}(p_4)\gamma^\nu \gamma_5(g_{\mu\nu}-\frac{p_{4\mu}p_{4\nu}}{m^2_4})\Big] \nonumber\\
&&\times \Big(p_2\!\!\!\!\!\slash+m_2\Big) \Big[ \Big(A_1 \gamma_\alpha \gamma_5+A_2 p_{2\alpha}\gamma_5+B_1 \gamma_\alpha+B_2 p_{2\alpha} \Big) \nonumber\\
&&\times u(p) \Big] \Big[-g_{KD^\ast D^\ast_s}\varepsilon_{\rho\sigma\tau\lambda}iq^\rho (-i)p^\tau_1\Big] \frac{-g^{\alpha\lambda}+p_1^{\alpha}p_1^\lambda/m^2_1}{p^2_1-m^2_1}\nonumber\\
&&\times \frac{1}{p^2_2-m^2_2}\frac{-g^{\mu\sigma}+q^\mu q^\sigma/m^2_E}{q^2-m^2_E}\mathcal{F}(q^2,m_{E}^2). \nonumber\\
\end{eqnarray}

In the same way, with the assumption that the $J^P$ quantum numbers of $P_{cs}$ are $3/2^-$, the amplitudes corresponding to diagrams (b) and (c) in Fig.~\ref{Fig:Tri} could be derived as follows:
\begin{eqnarray}
\mathcal{M}_b^{3/2}&=&i^3 \int\frac{d^4 q}{(2\pi)^4} \Big[g_{p_{cs}}^{3/2^-}\bar{u}_\mu(p_4) \Big] \Big(p_2\!\!\!\!\!\slash+m_2\Big)\Big[i(A+B\gamma_5)u(p) \Big]\nonumber\\
&&\times \Big[ig_{KD^\ast D_s}(ip_{3\nu}+ip_{1\nu}) \Big] \frac{1}{p^2_1-m^2_1}\frac{1}{p^2_2-m^2_2}\nonumber\\
&&\times \frac{-g^{\mu\nu}+q^\mu q^\nu/m^2_E}{q^2-m^2_E}\mathcal{F}(q^2,m_{E}^2).\nonumber\\
\mathcal{M}_c^{3/2}&=&i^3 \int\frac{d^4 q}{(2\pi)^4} \Big[g_{p_{cs}}^{3/2^-}\bar{u}_\mu(p_4) \Big] \Big(p_2\!\!\!\!\!\slash+m_2\Big) \nonumber\\
&&\times\Big[(A_1 \gamma_\nu \gamma_5+A_2 p_{2\nu}\gamma_5+B_1 \gamma_\nu+B_2 p_{2\nu})u(p) \Big]\nonumber\\
&&\times \Big[-g_{KD^\ast D^\ast_s}\varepsilon_{\rho\sigma\tau\lambda}iq^\rho(-i) p^\tau_1 \Big] \frac{-g^{\nu\lambda}+p_1^{\nu}p_1^\lambda/m^2_1}{p^2_1-m^2_1}\nonumber\\
&&\times \frac{1}{p^2_2-m^2_2}\frac{-g^{\mu\sigma}+q^\mu q^\sigma/m^2_E}{q^2-m^2_E}\mathcal{F}(q^2,m_{E}^2). \nonumber\\
\end{eqnarray}

In order to depict the structure and off shell effects, a  form factor in monopole form is introduced ~\cite{Cheng:2004ru, Tornqvist:1993vu, Tornqvist:1993ng, Locher:1993cc, Li:1996yn},
\begin{eqnarray}
\mathcal{F}(q^2,m^2) =\frac{m^2 -\Lambda_E^2}{q^2-\Lambda_E^2},\label{Eq:A1}
\end{eqnarray}
where the parameter $\Lambda_E$ is reparameterized as $\Lambda_E=m_{D^{\ast}}+\alpha \Lambda_{QCD}$ with $\Lambda_{QCD}=220$ MeV; and $m_{D^{\ast}}$ is the mass of the exchanged $D^\ast$ meson. The model parameter $\alpha$ should be of the order of unity~\cite{Tornqvist:1993vu, Tornqvist:1993ng,Locher:1993cc,Li:1996yn}. However, its precise value cannot be estimated by the first principle. In practice, the value of $\alpha$ is usually determined by comparing theoretical estimations with the corresponding experimental measurements.

With the above amplitudes, the partial width of $\Xi_b\rightarrow P_{cs} K$ could be estimated by
\begin{eqnarray}
\Gamma_{\Xi_b} &=& \frac{1}{2}\frac{1}{8\pi} \frac{|\vec{p}|}{m^2}\overline{\big|\mathcal{M} \big|^2},
\end{eqnarray}
where the factor $1/2$ results from the average of $\Xi_b$ spin and $\vec{p}$ is the momentum of $P_{cs}$ or $K$ in the rest frame of $\Xi_b$.  The overline indicates the sum over the spins of final states.

\section{Numerical Results and discussion}
\label{Sec:Num}
\subsection{Coupling Constants and Transition Form Factors}
The coupling constants relevant to the molecular states  can be estimated by the compositeness condition with the assumption that the observed $P_{cs}$ is in a molecule state which gives \cite{Xiao:2020Pre}
\begin{eqnarray}
g^{1/2^-}_{P_{cs}}=1.61^{+0.13}_{-0.08}, \ \ \  \, g^{3/2^-}_{P_{cs}}=2.82^{+0.20}_{-0.14},
\end{eqnarray}
where the center values are estimated with the model parameter $\Lambda=1.0 \ \mathrm{GeV}$, while the uncertainties of the coupling constants are obtained by varying the parameter $\Lambda$ from $0.8\ \mathrm{GeV}$ to $1.2\ \mathrm{GeV}$. The parameter $\Lambda$ in Ref.\cite{Xiao:2020Pre} comes from an exponent form of form factor, i.e. $\mathrm{exp}(p^2/\Lambda^2)$. It depicts the componential distribution in the molecular states, and it also plays the role of removing the ultraviolet divergences \cite{Xiao:2019mvs, Chen:2015igx}.

As an $S$-wave shallow bound state, the coupling constants of $P_{cs}$ and its components $ \Xi_c \bar{D}^\ast$ could be estimated under non-relativistic conditions~\cite{Weinberg:1965zz,Baru:2003qq}
\begin{eqnarray}
	g^2 =\frac{4\pi}{4m_0 m_2} \frac{(m_1+m_2)^{5/2}}{(m_1m_2)^{1/2}} \sqrt{32 E_b},
	\label{Eq:CP-Non}
\end{eqnarray}
where $m_0$, $m_1$ and $m_2$ are the masses of $P_{cs}$, $D^\ast$ and $\Xi_c$ respectively, and $E_b=m_1+m_2-m_0$ is the binding energy of the $S-$wave shallow bound state. With the measured mass of $P_{cs}$, the coupling constant is estimated to be $g=2.00$, which is very close to the estimation under compositeness conditions. Moreover, the coupling constants are also similar to those of $g_{P_c(4440) \bar{D}^\ast \Sigma_c} $ and $g_{P_c(4457) \bar{D}^\ast \Sigma_c}$ \cite{Xiao:2019mvs, Wu:2019rog}, which is consistent with the expectation of $\mathrm{SU}(3)$ symmetry.

The transition form factors of $\Xi_b\to \Xi_c$ could be parameterized in the form of \cite{Cheng:1996cs}
\begin{eqnarray}
f(Q^2)=\frac{f(0)}{(1-Q^2/m^2_V)^2},\ \ \ g(Q^2)=\frac{g(0)}{(1-Q^2/m^2_A)^2},
\label{Eq:A2}
\end{eqnarray}
where $m_V (m_A)$ is the pole mass of the vector (axial-vector) meson. For the $b\rightarrow c$ transition, the pole mass are $m_V=6.34$ GeV and $m_A=6.73$ GeV~\cite{Cheng:1996cs}. In Table. \ref{Tab:FFs}, the parameters related to the transition form factors of $\Xi_b\to \Xi_c$ \cite{Cheng:1996cs} are collected. It should be noticed that considering the form factor in Eq. (\ref{Eq:A1}) and the transition form factors, the loop integrals contain no ultraviolet divergences.

\begin{table}[t]
\begin{center}
\caption{The values of the parameters $f_i(0)$ and $g_i(0)$ in the form factors of $\Xi_b\to \Xi_c$ transition\cite{Cheng:1996cs}.}\label{Tab:FFs}
  \setlength{\tabcolsep}{2.4mm}{
\begin{tabular}{ccccccc}
  \toprule[1pt]
 ~~~ Parameter~~~ & ~~~ Value ~~~& ~~~Parameter~~~ & ~~~Value~~~\\
  \midrule[1pt]
                   $f_1(0)$ & $f_2(0)m_{\Xi_b}$  & $g_1(0)$ & $g_2(0)m_{\Xi_b}$  \\

                   0.533    & -0.124   & 0.580    & -0.019   \\
  \bottomrule[1pt]
\end{tabular}}
\end{center}
\end{table}

\subsection{Partial Width of $\Xi_b \to P_{cs} K$}

\begin{figure}[htb]
  \centering
 \includegraphics[width=1.0\linewidth]{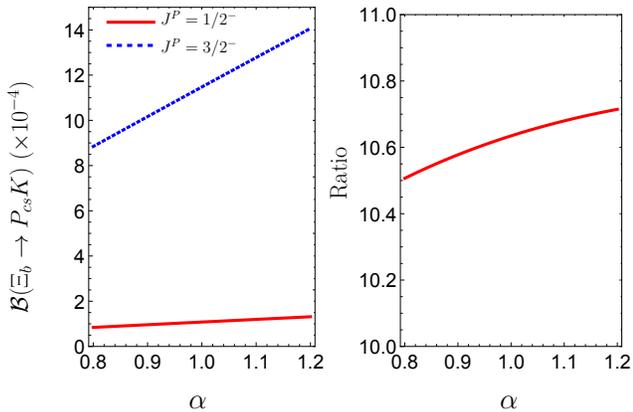}
  \caption{The branching fractions of  $\Xi_b\rightarrow P_{cs}K$ with different $J^P$ assignments (left panel) and the ratio of the branching fractions (right panel) depending on the parameter $\alpha$.}\label{Fig:Br}
\end{figure}

To date, the $J^P$ quantum numbers of $P_{cs}$ have not been determined by LHCb collaboration. In the $S$-wave $\bar{D}^\ast \Xi_c$ molecular scenario, the possible assignments can be $J^P=1/2^-$ and $3/2^-$. In the present work, the branching ratios of $\Xi_b \to P_{cs} K$ with two different $J^P$ assignments are estimated. The model parameter $\alpha$ varies from 0.8  to 1.2, which is the same as the one of $P_c$ production~\cite{Wu:2019rog}. In this $\alpha$ range, the branching ratio is estimated to be $(0.84\sim1.31)\times 10^{-4}$ for $J^P=1/2^-$ and $(8.85\sim14.1)\times 10^{-4}$ for $J^P=3/2^-$. It indicates that the branching ratio of $\Xi_b \to P_{cs} K$ should be of the order of $10^{-4}$. Moreover, the $\alpha$ dependences for two different assignments are found to be very similar, the ratio of the branching fractions with different $J^P $ assignments is presented in the right panel of Fig.~\ref{Fig:Br}. The branching ratio with $J^P=3/2^-$ is about one order of magnitudes larger than that with $J^P=1/2^-$. Our estimation indicates that the ratio is around 10.6, and its dependence on the model parameter is weak. Thus, a molecular state of $\bar{D}^\ast \Xi_c$ with $J^P=3/2^-$ is more simply produced from $\Xi_b$ decay.

In Ref.~\cite{Wu:2019rog}, the authors argued that the production of $P_c$ states occurs via $\Lambda_b \to \Sigma_c D_s^{(\ast)} \to P_{c} K$. However, the weak transition of $\Lambda_b \to \Sigma_c$ should be suppressed with a suppression factor $R=0.09$ due to isospin violation and spin flip of the light di-quark. In Refs.~\cite{Liu:2015fea, Du:2021fmf}, the authors indicated that there can be important contribution from other color suppressed diagrams. The present estimation indicate that the branching fraction of $\Xi_b \to P_{cs} K$ occur via a very similar process but without isospin violation and spin flip of the light di-quark. The branching ratio is of order of $10^{-4}$, which is about two orders larger than that of $\Lambda_b \to P_{c} K$.

Moreover, the experimental analysis of LHCb collaboration indicated that the fit fraction of $P_{cs}$ is $(2.7^{+1.9+0.7}_{-0.6-1.3})\%$, which indicates~\cite{Aaij:2020gdg}
\begin{eqnarray}
\frac{\mathcal{B}\left(\Xi_b \to P_{cs} K \to J/\psi \Lambda K \right)}{\mathcal{B}\left(\Xi_b \to J/\psi \Lambda K\right)}=\left(2.7^{+1.9+0.7}_{-0.6-1.3}\right)\%, \label{Eq:BR1}
\end{eqnarray}
while the branching fraction of $\Xi_b \to J/\psi \Lambda K$ is measured to be~\cite{Aaij:2017bef}
\begin{eqnarray}
	\frac{f_{\Xi_b}}{f_{\Lambda_b}} \frac{\mathcal{B}(\Xi_b \to J/\psi \Lambda K)}{\mathcal{B}(\Lambda_b \to J/\psi \Lambda)} = \left( 4.19 \pm 0.29 \pm 0.15\right) \times 10^{-2}, \label{Eq:BR2}
\end{eqnarray}
with $f_{\Xi_b}$ and $f_{\Lambda_b}$ being the fragmentation fractions of $b \to \Xi_b$ and $b\to\Lambda_b$ transitions respectively. The branching fraction of $\Lambda_b \to J/\psi \Lambda$ has been measured by CDF collaboration, which is~\cite{Abe:1996tr}
\begin{eqnarray}
	\mathcal{B}(\Lambda_b \to J/\psi \Lambda)=(3.7\pm 1.7 \pm 0.7) \times 10^{-4}. \label{Eq:BR3}
\end{eqnarray}
This measured branching fraction is consistent with the theoretical estimation in Ref.~\cite{Cheng:1995fe}, which is $2.1 \times 10^{-4}$. As for the fragmentation fractions, the theoretical estimations are model dependent~\cite{Cheng:2015ckx, Hsiao:2015txa, Voloshin:2015xxa, Jiang:2018iqa, Wang:2019dls}. Using the data sample at $\sqrt{s}=7,\ 8 $TeV, the recent experimental analysis from the LHCb collaboration shows that the ratio of the fragmentation fractions is~\cite{Aaij:2019ezy},
 \begin{eqnarray}
 	\frac{f_{\Xi_b}}{f_{\Lambda_b}} =(6.7\pm 0.5 \pm 0.5  \pm 2.0) \times 10^{-2},  \label{Eq:BR4}
 \end{eqnarray}
which is very consistent with the estimations in Refs. \cite{Jiang:2018iqa, Wang:2019dls}.  With the above experimental results, the branching ratio of $\Xi_b \to P_{cs} K \to J/\psi \Lambda K$ could be obtained as follows:
 \begin{eqnarray}
 	\mathcal{B}(\Xi_b \to P_{cs} K \to J/\psi \Lambda K)=\left( 6.25^{+5.98}_{-4.98} \right) \times 10^{-6}.
 \end{eqnarray}

Considering the small width of $P_{cs}$, an approximation of $\mathcal{B}(\Xi_b \to P_{cs} K)\times \mathcal{B}(P_{cs} \to J/\psi \Lambda) \simeq \mathcal{B}(\Xi_b \to P_{cs} K \to J/\psi \Lambda K)$ could be obtained. The present estimations indicate the branching ratio of $\Xi_b \to P_{cs} K$ is of the order of $10^{-4}$, thus the expected branching ratio of $P_{cs} \to J/\psi \Lambda$ should be of the order of $10^{-2}\sim 10^{-1}$. In Ref.~\cite{Shen:2019evi}, bound states of $\bar{D}^\ast \Xi_c$ are found and the branching fraction of $J/\psi \Lambda$ mode is estimated to be of the order of $10^{-2}$ for $J^P=1/2^-$ and $10^{-1}$ for $J^P=3/2^-$, which is consistent with the above analysis. In addition, in the estimations of Ref.~\cite{Shen:2019evi}, the dominant decay modes of $P_{cs}$ state are $\rho \Sigma$ and $K^\ast \Xi$.

Before the end of this work, it is worth mentioning that the two resonances have been fitted by the LHCb collaboration~\cite{Aaij:2020gdg}. However, the analysis indicate that the two-peak hypothesis cannot be confirmed due to limit of the current data. In addition, in the two resonance fit scheme, the masses of the resonances are fitted to be $4454.9 \pm 2.7$ MeV and $4467.8 \pm 3.7$ MeV, respectively, which are consistent with that of $P_{cs}(4459)$ within uncertainty even when the systematically uncertainties are excluded in the two-resonance scenario. Experimentally, more precise data is expected in future, which would provide more information on the charmed-strange pentaquark states.

\section{Summary}
\label{Sec:Summary}
In summary, we have presented an investigation on the production of the newly observed $P_{cs}$ state from $\Xi_b$ decay in a $\bar{D}^\ast \Xi_c$ molecular scenario. By analyzing the $P_{cs}$ production at quark level, the dominant contributions to $P_{cs}$ productions are found to be  the charmed-strange hadron loops, where the initial $\Xi_b$ couples to $\Xi_c \bar{D}^{(\ast)}_s$ by weak interaction, and then the $\bar{D}^{(\ast)}_s$ transits into $\bar{D}^\ast$ via kaon emission, and the recoil $\bar{D}^\ast$ and $\Xi_c$ form the molecular state, i.e., $P_{cs}$.

The production process of $\Xi_b\rightarrow P_{cs}K$ is estimated  at hadron level with an effective Lagrangian approach. Our estimation indicates that the branching fraction of $\Xi_b \to P_{cs} K$ is of the order of $10^{-4}$ for both $J^P=1/2^-$ and $3/2^-$. Moreover, we find the ratio of the branching fractions with $J^P=3/2^-$ and $1/2^-$ is about 10.6 and almost independent of the model parameter. Our analysis indicates that the estimated productions fraction in the present work are consistent with the experimental measurements and the theoretical estimations in the literature available.

\section{ACKNOWLEDGMENTS} D. Y. Chen would like to thank Professor Wei Wang and Professor Fu-Sheng Yu for useful discussions. This work is supported by the National Natural Science Foundation of China under the Grant No. 11775050. Qi Wu was also supported by the Scientific Research Foundation of Graduate School of Southeast University (Grants no. YBPY2028).

\end{document}